DTIP of MEMS & MOEMS 9-11 April 2008# Measurement of Large Forces and Deflections in Microstructures

Kai Axel Hals[1], Einar Halvorsen, and Xuyuan Chen
Institute for Microsystem Technology, Vestfold University College,
P.O. Box 2243, N-3103 Tønsberg, Norway
*Abstract*-Properties of typical MEMS materials have been widely investigated. Mechanical properties of MEMS structures depend not only on the bulk material properties, but also structural factors. A measurement system has been made to measure force/deflection on microstructures to examine some of the structural properties. This is a stylus setup integrated with a load cell and a linear actuator. First, the requirements for the measurement system were established. Then the system was built up and characterized. We have successfully made measurements on a typical micromechanical structure, a cantilever accelerometer design. The stylus placement accuracy, the spring constant along the proof mass, analysis of the force/deflection curve shape and destructive tests on the cantilever have been investigated in our experiment and will be presented in this paper.
I. INTRODUCTION

The direct wafer level measurement of force versus deflection of MEMS structural elements is not available in conventional probe stations, but is highly desired. Verification of the elastic stiffness at specific points could be useful as design verification and possibly also in process control. Furthermore, if sufficiently large deflections could be made, it would allow for strength assessment at wafer level. For accelerometers this would be an alternative to shock testing of packaged devices or could serve as a complementary test.

Measurement of displacement versus force can be done by atomic force microscopy, but is limited with respect to displacement and scan range. Surface profilometers could in principle also be used for this purpose [1, 2], but are limited with respect to force range and are not suitable for measurements at a point. There are nano-indenters that have both the displacement and the scan range necessary to make them useful for this kind of measurements [3, 4]. These expensive instruments are optimized towards characterization of thin films and surfaces, not for test of MEMS structural elements. In particular they have much finer spatial resolution and different tip geometry than needed for measurement of large deflections of a structural element such as a beam or a suspended proof mass.

From the above, it is clear that there is a need to investigate methods for mechanical test of MEMS structural elements directly on wafer. It is therefore worthwhile to investigate if a simple and affordable approach for mechanical testing of MEMS can be found. In particular it is interesting to investigate large deflection for potential fracture test on wafer.

We have built a simple and affordable measurement setup for probing mechanical properties of MEMS structures at wafer level. In the following we give a detail description of our setup and analyze its behaviour over a wide range of loads based on measurements on micromachined silicon accelerometers.

II. MEASUREMENT SETUP

The basic principle of the apparatus is shown in Fig. 1. A probe tip is attached to a load cell which measures the vertical force on the probe tip. The load cell is attached to a linear z-actuator which is computer controlled with known displacement. The above equipment is mounted on a fixture that is attached to a manual xy-actuator (a small xy-table). Since the z-displacement is known and the force is measured, we can obtain the force versus deflection response at any point of choice on a device.

The measurement setup is depicted in Fig. 2a. The load cell was a Honeywell 25g Minigram beam load cell, the linear actuator was a Zaber CE Linear actuator. The actuator and the load cell were connected to a PC. The actuator was connected through the RS-232 interface and the load cell was read by a Texas Instruments USB 6009 ADC, both controlled by a LabView program. In addition a laboratory microscope was used for visual inspection and probe tip positioning. The whole setup was placed on an anti-vibration table.

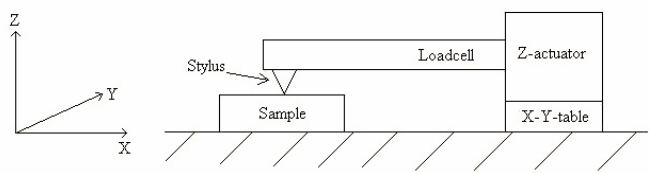

Fig. 1. Basic principle of measurement setup.

[1] Present address: Infineon Technologies Sensonor AS, P.O.Box 196, N-3192 Horten, Norway
©EDA Publishing/DTIP 2008    ISBN: 978-2-35500-006-5

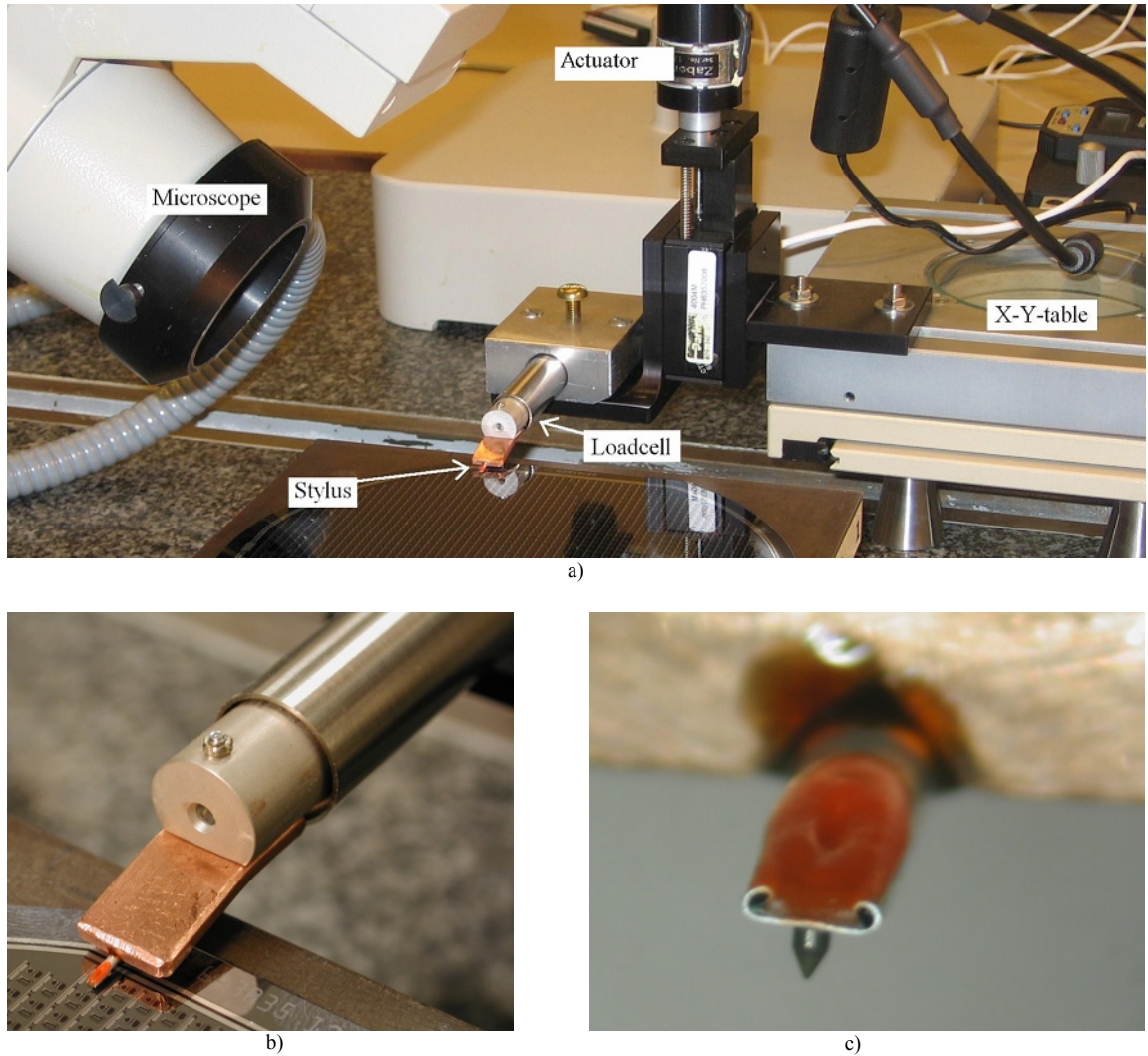

Fig. 2. Measurement setup realisation: a) overview, b) probe tip and load cell, and c) probe tip.

The probe tip is a very critical element. In our initial design considerations for the setup, we analyzed the maximum stress in the tip based on the Hertz contact problem. For a tip radius of 10um, a silicon substrate and 10mN tip force, the maximum stress is several GPa for several choices of tip material. We therefore opted for a diamond tip. The probe was made by removing the cantilever from a Shure SS35C-Q stylus and gluing the cantilever into an alignment groove in a copper plate. The copper plate was fixed to the load cell by a screw as depicted in Fig. 2b. The SS35C-Q stylus needle is made from polished natural diamond and has a tip spherical radius of 18um [5]. A close-up of the probe tip is shown in Fig. 2c.

The accuracy in the vertical displacement is given by the precision of the linear actuator. Even though the step-size is about 0.1um, there is a systematic cyclic error in the positioning of +/-1um according to the manufacturer. For large displacements, we can safely neglect this error. For stiffness estimates the systematic error can be cancelled by sampling at the spatial period of the systematic error. This is sufficient for our purpose.

Without any precautions, the dominating error in the force measurements comes from the electrical noise in the load cell. By estimating the power spectral density of the electrical signal from the load cell, we have verified that it is essentially flat so that the correlation time is correspondingly very short. The standard deviation in the readout is about 1mN when estimated from a long time series. This is far too much for our requirements. To increase accuracy we have adopted a sampling frequency of 20kHz and used the average readout over 3 seconds at this sampling rate for the estimate of the force. This procedure reduces the standard deviation in the estimates to 4.4uN which is sufficient.

The mounting of the probe tip explained above, gives a slightly different loading of the load cell than what it is calibrated for. In order to account for this, we have verified and calibrated the setup by measuring the force when pushing the tip onto the scale pan of a precision micro scale and comparing to the readout of the scale. This test shows that the mounting gives a systematic overestimate of the force of about 2.5% which we correct for.

The scale pan is based on a force restoration principle so that it does not deflect when loaded. By assuming that the displacement that is set for the actuator is completely accounted for by the deformation of the apparatus itself, we can estimate the stiffness of our measurement system. We find 4635N/m. This is one to two orders of magnitude stiffer than the structures that we have measured.

### III. MEASUREMENTS

To investigate the capabilities of our setup, we have performed measurements on an accelerometer design that were fabricated in a silicon bulk micromachining process with beam and proof mass thicknesses defined by etch stop on pn-junction, a cantilever type design. This design had a symmetrically placed single beam and the nominal stiffness when loaded by a point force at the centre of mass is about 10N/m. All measurements reported here, were made by placing the probe tip on top of the proof mass such that the forces act in the plane of symmetry of the device.

Measurements were made by placing the probe tip initially at the symmetry centre of the top surface of the proof mass, see Fig 3a. Then the probe tip was displaced vertically (orthogonal to the wafer plane) in controlled steps over a range of deflections from 0 to about 300um. At each displacement the force was measured. The load cell is sensitive in the vertical direction. Hence it is only that component of the force that is read out from the apparatus. The resulting force versus deflection curve is shown in Fig. 4.

Roughly we can divide the force readout into three characteristic regimes as indicated in the Fig. 4: 1) a linear regime for deflections up to about 50um, 2) a regime of geometric nonlinearities from about 50um to 230um, and 3) a regime with contact nonlinearities from about 230um to fracture.

#### A. Stiffness in the linear regime

When the proof mass deflection is sufficiently small that the loading conditions are not substantially altered, we have the linear regime. The measured force then equals the actual force between the tip and structure and is directed in the vertical direction as indicated in Fig. 3a. In this case the force readout is well described by Hookes law $F_z=kz$, where $k$ is the stiffness of the structure at the point where the load is applied and $z$ is the vertical displacement of the point.

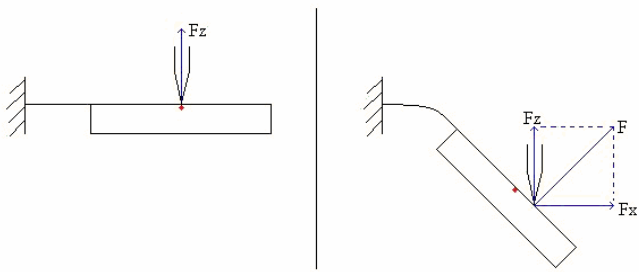

Fig. 3. Load, a) linear regime b) geometrically nonlinear regime.

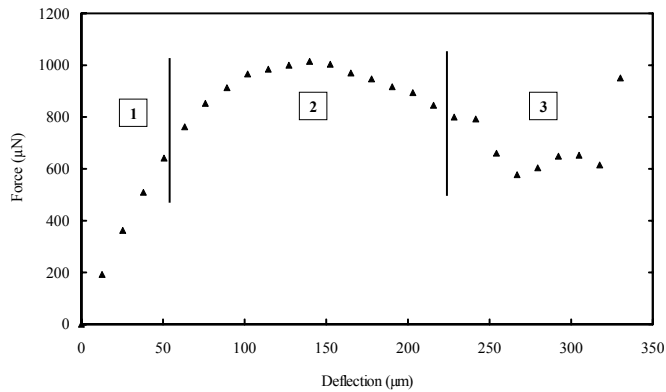

Fig. 4. Force readout versus deflection.

This stiffness will vary with the position of the load. We measured it as a function of position for various probing positions along the proof mass from one edge to the other. For each probing position, 3-4 measurements of force versus deflection over a small range of deflections well within the linear range were made. The stiffness was then estimated by a linear fit to the force versus deflection measurement at each position. The range of deflections in this experiment was from about 5um to 50um and the forces were in the range from 20uN to 1.3mN.

The resulting stiffness versus position trace is shown in Fig. 5 together with corresponding data obtained from analytical calculations (rigid mass and Euler-Bernoulli theory for the beam). In the figure position 0μm is at the support of the beam. The calculated results are in good qualitative agreement with the measurement, but systematically overestimate the value of the stiffness. We attribute this to the neglect of support compliance in the calculated results.

#### B. Geometric nonlinearities

When the proof mass is sufficiently displaced and rotated the loading conditions are significantly altered and we have the regime of geometric nonlinearities. This situation is shown in Fig. 3b. Assuming negligible friction between the probe tip and proof mass, the force $F$ on the proof mass must be orthogonal to the surface. This force can be decomposed into a component in the vertical direction $F_z$ and a component in the horisontal direction $F_x$. The sensitive direction of the load cell is in the vertical direction, so it is only $F_z$ that is measured. As the deflection increases, so does the rotation of the proof mass and $F_z$ becomes a correspondingly smaller fraction of the total force.

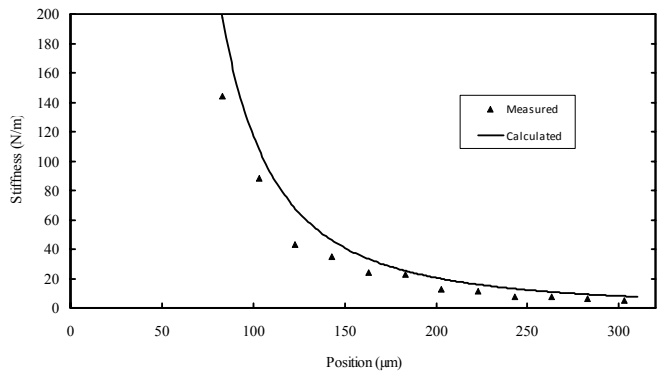

Fig. 5. Spring constant versus position along proof mass.

The point of contact slides towards the outer edge of the proof mass as the deflection grows. Therefore, for a given total force $F$, the bending moment in the beam will be larger than it would have been if the point load had stayed fixed with respect to the surface. Due to the larger bending moment in the beam, the $F$ necessary to sustain a certain rotation of the proof mass is then smaller than it otherwise would have been.

The net result is that the measured force component $F_z$ initially increases, reaches a maximum, and then decreases from a deflection of 140um onwards, even though total force $F$ increases.

There is a small effect that acts against this effect. The stylus is not perfectly sharp, but has a tip radius of 18 μm. This will cause a rolling effect that moves the point of the applied force inwards. This effect is limited from above by the tip radius which is small compared to the length that the tip slides on the mass.

*C. Contact nonlinearities and destructive tests*

We refer to regime no. 3 in Fig. 4 as the regime of contact nonlinearities. Here, the way that the probe tip makes contact with the proof mass changes. When the initial position is at the centre of symmetry, the main sequences of events are as depicted in Fig. 6. First the probe tip slides along the proof mass (top panel), then it slides off the proof mass edge (middle panel), then it slides on the edge until finally the proof mass surface becomes parallel to the tip surface and there is distributed contact between the tip and the proof mass surface (bottom panel).

The small "bump" seen in the leftmost part of the contact nonlinearity regime in Fig. 4, is due to the tip sliding off the edge. The following irregular behavior is due to the tip sliding on the edge, then making distributed contact with proof mass and finally, for sufficiently large deflections, the region of contact extends onto the beam. In the latter case, the force distribution becomes very difficult to analyze. It is clear that the structure is substantially stiffer when loaded in this region. This is the reason why there is an increase in the measured force values in the end of the curve. When the structure is deflected further, it fractures.

It is possible to avoid that the tip slides off the mass by placing the probe tip further from the outer edge. We therefore did another experiment where the stylus is placed 50 μm closer to the spring than the centre of symmetry (COSYM). In Fig. 7 the result of this measurement is shown together with the original measurement for comparison.

The second measurement shows both a linear and a geometrically nonlinear regime similar to the first. The differences are due to the much higher stiffness which results in larger forces and a smaller range of deflections.

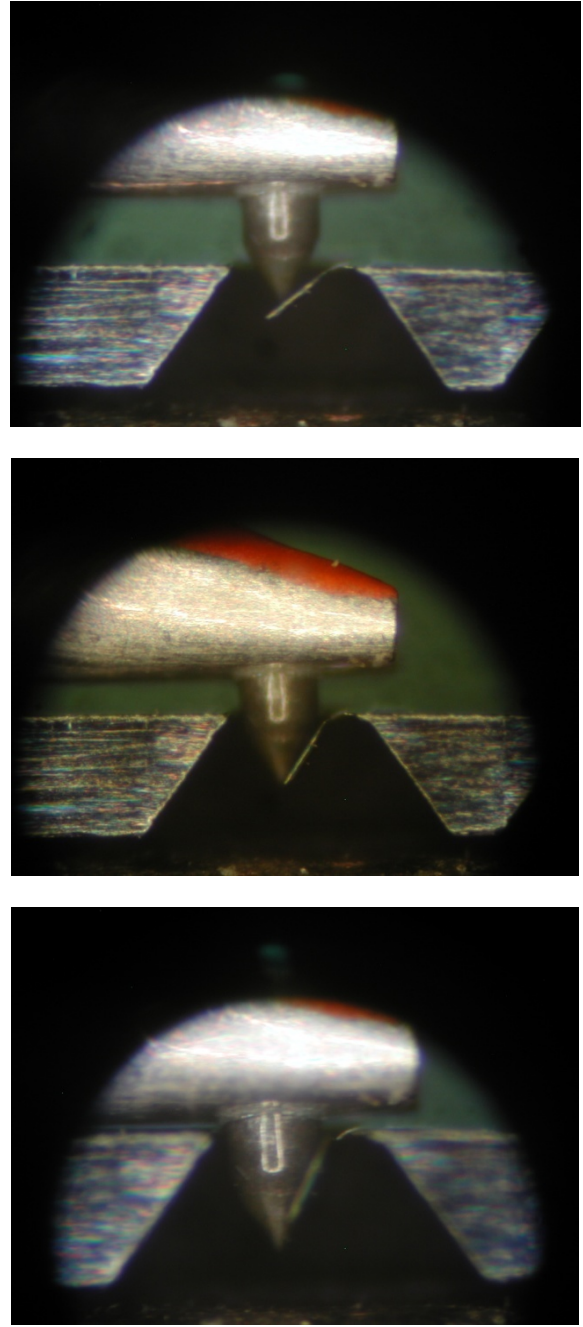

Fig. 6. Stylus at different loading conditions.

In the second measurement, the stylus does not slide off the mass. Also, for sufficiently large deflections, distributed contact is achieved, but extends onto the beam immediately. Hence, the irregular behavior of the first measurement is absent in the second. The sharp rise in force at large deflection is due to the tip being in distributed contact with beam in both cases.

Due to the rigidity of the proof mass, the distributed contact between the probe tip and proof mass is entirely dictated by the probe tip angle regardless of the initial probe tip position. It is possible to avoid the distributed contact by tilting the probe or using a probe tip with one vertical side, but unfortunately that is not an option in our current setup.

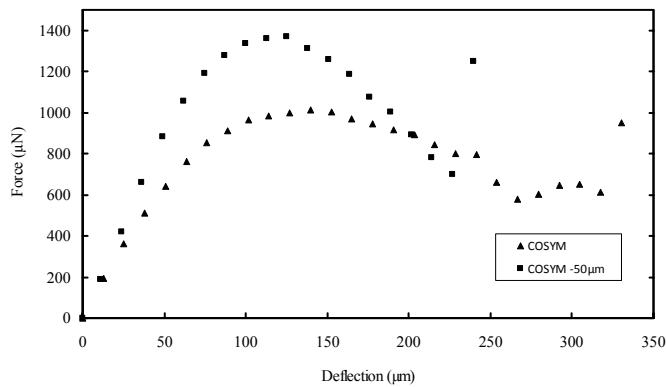

Fig. 7. Force readout for different initial placements of stylus.

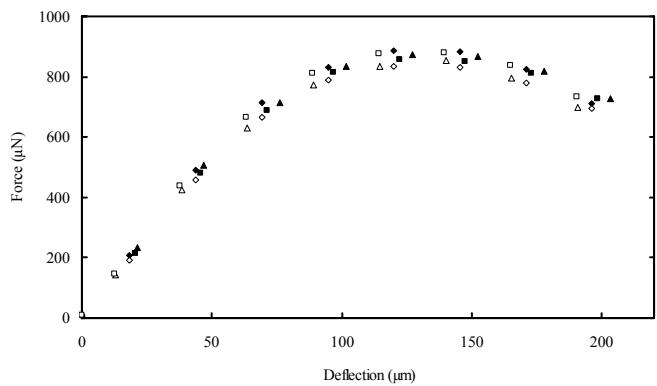

Fig. 9. Sample to sample variations on wafer.

In order to estimate fracture strength, we need to have a well defined loading of the device. The distributed contact prohibits that for the tested design because it is compliant enough to reach a slope equal to the probe tip angle before fracture.

To investigate a situation without the distributed contact, we measured samples of a stiffer accelerometer design. This design had two beams placed widely apart and a nominal stiffness at the centre of mass of about 60N/m. It is described in [6], and the measurement is shown in Fig. 8. In this case fracture arises before the distributed contact took place, but the probe tip just barely slid off the edge prior to fracture of the beam. Both devices fractured at about 4.5mN vertical force and about 110um deflection. The last point on the curve before the tip slid off was 4.0mN at 90um deflection.

*D. Sample to sample variations*

To further investigate the capabilities of the measurement setup, we have made a series of measurements on wafer. Fig. 9 shows the test results measured at the centre of the mass of six accelerometers located at different sites on the wafer. As seen from Fig. 9, there exist some variations in the six measurements.

From the setup characterization processes, we found that the standard deviation of one point in a measurement was found to be 4.4μN. In addition the accuracy of the placement of the stylus will play a significant role in determining the accuracy of the measurement. With the current measurement setup, where the stylus is placed optically through a microscope, it is difficult to place the stylus on the exact same place on each structure. By placing the stylus in a line prolonging the cantilever, it was found that a deviation of placement of the stylus of +/10μm would cause a variation of up to +/- 60μN in the force versus deflection curve. Though above mentioned factors contribute to the uncertainty in the measurements, the deviation in placement of the stylus dominates. The variations presented in Fig. 9 are well within the range caused by the placement inaccuracy. The variation from accelerometer to accelerometer across the wafer are simply too small to detect with this measurement setup. With the possibility to place the stylus more accurately, the main reason for the measurement variations would disappear. This could be done with a more precise x-y-table, or a more suitable optical instrument for placement.

IV. CONCLUSIONS

In conclusion, we have designed, built and characterized a system for wafer level force versus deflection measurements on microsystems. With an accuracy of 1um in displacement and 4.4uN resolution in force up to a range of 250mN, we have sufficient precision to measure stiffness at chosen points on the device. The mechanical behavior of the test accelerometer measured with the system, from small to large deflections of the cantilever beam, all the way up to the fracture of the beam, was analyzed and understood. Measurements on two types of accelerometers provide a solid basis for future development.

ACKNOWLEDGMENT

Infineon Technologies SensoNor AS is acknowledged for supplying test structures and financial support. We appreciate Dr. Trond Inge Westgaard for his initiative of building this setup and discussions during realization of the setup.

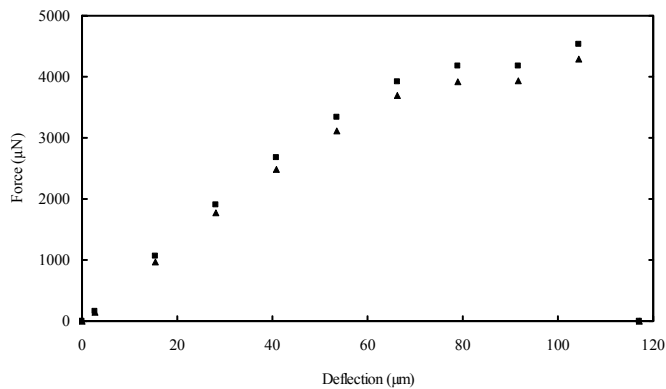

Fig. 8. Force versus deflection readouts from a stiffer accelerometer design.

REFERENCES

[1] M. W. Denhoff, "A measurement of Young's modulus and residual stress in MEMS bridges using a surface profiler," *Journal of Micromechanics and Microengineering*, vol. 13, pp. 686-692, 2003.

[2] M. Hopcroft, T. Kramer, G. Kim, K. Takashima, Y. Higo, D. Moore, and J. Brugger, "Micromechanical testing of SU-8 cantilevers," *Fatigue & Fracture of Engineering Materials & Structures*, vol. 28, pp. 735-742, 2005.

[3] O. Kraft and C. A. Volkert, "Mechanical testing of thin films and small structures," *Advanced Engineering Materials*, vol. 3, pp. 99-110, 2001.

[4] G. M. Pharr and O. W. C., "Measurement of thin film mechanical properties using nanoindentation," *MRS Bulletin*, vol. 17, pp. 28-33, 1992.

[5] User Guide Shure SS35C-Q, . http://www.shure.com/ProAudio/Products/Accessories/us_pro_SS35C_content

[6] C. Lowrie, C. Grinde, M. Desmulliez and L. Hoff. "Piezoresistive three-axis accelerometer for monitoring heart wall motions", *in Procs. DTIP*, Tima Labs, Grenoble, 2005, pp. 131-136.